# Molecular Identification, Antioxidant Efficacy of Phenolic Compounds, and Antimicrobial Activity of Beta-Carotene Isolated from Fruiting Bodies of Suillus sp

Shimal Yonuis Abdulhadi
*University of Mosul*, shimal_y@yahoo.com

Raghad Nawaf Gergees
*University of Mosul*, raghadnawaf@yahoo.com

Ghazwan Qasim Hasan
*University of Mosul*, dr.ghazwan@uomosul.edu.iq

# Molecular Identification, Antioxidant Efficacy of Phenolic Compounds, and Antimicrobial Activity of Beta-Carotene Isolated from Fruiting Bodies of Suillus sp

## Abstract

*Suillus* species, in general, are edible mushrooms, and environmentally important that are associated mostly with pine trees in the tropics regions. These fungi considered a remarkable source of phenolic compounds that play a crucial role as antioxidants which may reduce the risk of most human chronic diseases such as cancer, diabetes, asthma, atherosclerosis, Alzheimer, and others. On the other hand, carotenoids (*β* carotene) are the most popular natural pigments which play an important role to protect the plants from photo-oxidative reactions. In human, these compounds prevent oxidative stress and expects to have antimicrobial activity. Here, the phenolic compounds were extracted with Ethyl acetate from fruiting bodies of *Suillus* sp and analyzed by HPLC, the antioxidant activity (reducing power%) of phenolic compounds was determined at the concentrations of 1, 2.5, and 5 mg/mL. Antimicrobial activity of *β* carotene pigment was measured at a concentration of 100 mg/mL against some human pathogenic bacteria such as *Escherichia coli, Pseudomonas aeruginosa, Klebsiella pneumonia*, and *Staphylococcus aureus*. The specific DNA region ITS was amplified and sequenced using ITS1 and ITS4 primers with some bioinformatics analyses. The phenolic extract isolated from fruiting bodies of *Suillus* sp showed a remarkable antioxidant activity by increasing the reducing power percent (from $F^{+3}$ ions to $F^{+2}$ ions) comparing with the industrial antioxidant (Propyl gallate) at all used concentrations. Percent of reducing power of phenolic compounds were 75.5, 84.9 and 95.7% at concentrations of 1, 2.5, and 5 mg/mL respectively; comparing with PG were 65.9, 81.3, and 93.3 at 1, 2.5, and 5 mg/mL respectively. The *β* carotene pigment revealed a significant antimicrobial activity at a concentration of 100 mg/mL against *K. pneumonia, E. coli*, and *S. aureus*. The highest bacterial growth inhibition was against *K. pneumonia* (40 mm), followed by *E. coli* (36 mm) and *S. aureus* (31 mm), while no effect showed against *P. aeruginosa*. Our outcomes revealed that the phenolic bioactive compounds can be used as a natural antioxidant instead of the industrial antioxidants, and also a *β* carotene pigment could be applied as a promising natural compound rather than using the antibiotics and other manufactured compounds to inhibit bacteria activity.

## Keywords

Suillus sp, beta-Carotene, Fruiting bodies, Antioxidant activity, Antimicrobial activity.

## Cover Page Footnote

The authors are very grateful to the University of Mosul/ College of education for the pure sciences/ Department of Biology for their provided facilities, which helped to improve the quality of this work.

## 1. Introduction

*Suillus* species are common and ecologically important fungi, that form ectomycorrhizal relations at most with Pine trees [1−3]. *Suillus* members are mainly distributed in northern temperate regions [2,3]. Many studies have used the internal transcribed spacer (ITS) region for detection purposes as this region is renowned in *Suillus* [3,4] and many other fungi [5,6]. Sequencing of ITS genes has become a fundamental research approach to microorganisms classification [7]. Compared with a basic fungal culture, ITS sequencing is not limited to any environment as well as to sample size. In addition, ITS sequencing appears to be more cheaper with global availability than gene chipping, multi-locus sequencing or nanopore sequencing [8]. The researchers were interested in the field of natural products by extracting the active compounds from fruiting bodies of fungi, and studying the antimicrobial and antioxidant effects of these compounds against bacteria as well as fungi [9]. These natural compounds which contain many active substances were considered an important start for modern medicines and stimulated the interest of the companies specialized in manufacture of medicines, medical preparations, and cosmetics to use them as raw materials in their products. Both [10] have explained that antibacterials from natural products are highly effective in treating many infectious diseases without having side effects, that are often caused by chemically manufactured antibiotics. Another study has revealed the antimicrobial activity of ethanolic extract of *Leucoagaricus leucothites* in some foodborne and spoilage bacteria, were found the chelating effect on ferrous ions was 82% at15 mg/mL, while the scavenging effect on DPPH radicals was 71% at 10 mg/mL of ethanol extract [11]. The oxidation reactions are necessary for life but can be harmful and result in free radicals that start with series reactions that lead to damage the cells [12]. Therefore, one of the most important areas that attracted a lot of attention from researchers was to find the possible therapeutic potential of antioxidants in controlling oxidative diseases associated with oxidative damage; hence many natural extracts have been detected that have outstanding activity as an antioxidant [13]. Hayashi et al. [14] found the possibility of producing antioxidants from microbial sources, were found that *Pseudomonas roqueforti* mold can produce antioxidants when grown on liquid media; the 2,3-Dihydroxybenzoic acid compound was diagnosed as having antioxidant activity close to industrial antioxidant activity (Butylated hydroxyanisole) abbreviates as BHA. Nowadays, it becomes necessary to find antibiotics that have a new synthetic structure that differs from other antibiotics, due to microorganisms resistance development against antibiotics, as they have the ability to form a biofilm and increase antibiotic resistance [15]. Pigments have always applied against bacteria and fungi as a natural product compound. The antimicrobial activity of β-Carotene was showed in a bovine lactoperoxidase system of *Salmonella enteritidis* which increased from 1.4 to 3.8 log units through the addition of 20-fold diluted CE. β-Carotene, a major natural pigment of carrot [16]. In addition, phenols have been showed antioxidant effects, this is due to their abilities to protect organic compounds present in the body from oxidation reactions and protect the body from chronic diseases [17,18]. Studies have confirmed the benefits of natural antioxidants, and their impacts to reduce chronic diseases risks or prevent disease progression either by strengthening the body with natural antioxidant defenses or using it as a dietary supplement in the diet [19]. It has been proven that phenolic compounds have remarkable effectiveness in protecting against oxidation in various laboratory tests, as it works to scavenge harmful free radicals in the body, and it has used safely to solve skin problems [18]. The present study aims to evaluate the antioxidant activity of phenolic compounds extracted from fruiting bodies of *Suillus* sp, and the inhibitory activity of β-carotene pigment on different bacterial genera.

## 2. Materials and methods

### 2.1. Fungi samples collection

Fruiting bodies of *Suillus* sp were collected from farms located on the Tigris river/Al-Rashidiya quarter/Mosul city, north-west of Iraq in December 2019. The samples were transported to the lab in sterile polythene bags (Fig. 1).

### 2.2. Quantitative estimation of β-carotene and HPLC analysis

The method presented by Khatua et al. [20] was used to estimate the amount of beta-carotene in methanol extract, were 100 mg of methanol crude extract was dissolved in 10 mL of HPLC grade acetone-hexane solution (4:6), then the sample was filtered through Whatman No 4. Absorbance was detected simultaneously at 453, 505, and 663 nm. The amount of carotenoids is calculated using the formula below.

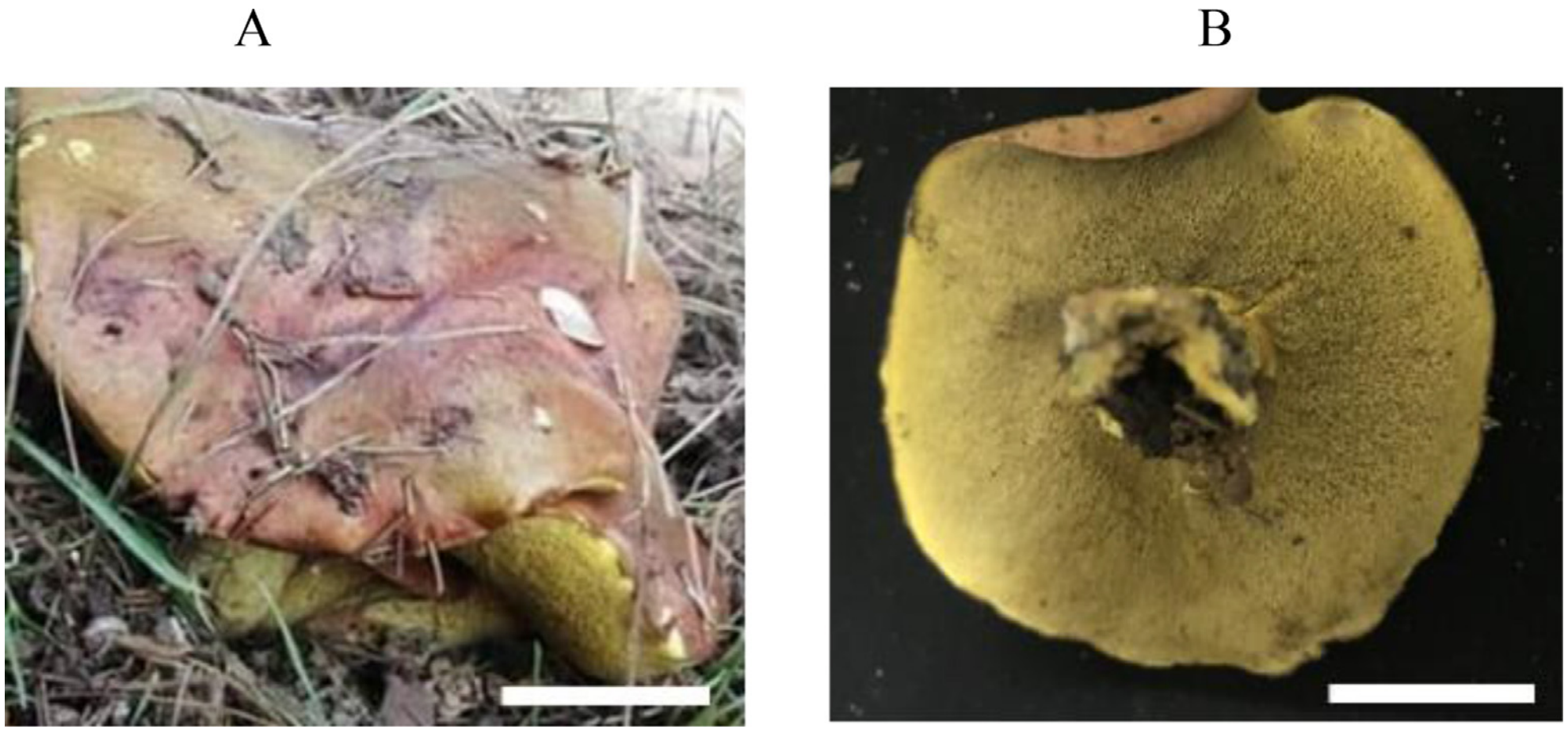

Fig. 1. Fruiting bodies of *Suillus* sp. (A) An image of the fungus taken from the field (B) The ventral side of the fungus. Scale bars = 2 cm in A and B.

$$\beta - \text{carotene}(\text{mg}/100\ \text{mL}) : 0.216\ A_{663} - 0.304\ A_{505} + 0.452\ A_{453}$$

The HPLC analysis was performed in the Ministry of Science and technology/Department of environment and water/Baghdad-Iraq. About 20 μl filtrate was loaded on the HPLC system model (SYKAM) Germany, the separation process was carried out at a flow rate of 0.8 mL/min at 25 °C. The mobile phase was (Acetonitrile: DW) (80: 20), the column is C18-ODS (25 cm * 4.6 mm) and the detector (UV–Vis) at a wavelength 450 nm at a flow rate 1 mL/min.

### 2.3. Extraction and antioxidant activity of phenolic compounds

About 10 g of powdered *Suillus* sp fruit bodies were soaked in 200 mL methanol for two days, the filtered residue was re-extracted with methanol solvent, the mixtures were evaporated at 40 °C under reduced pressure (Rotavapor R-3, Butchi, Switzerland) to obtain methanol extract of *Suillus* sp fruit bodies which contains the bioactive compounds. The reducing power of the phenolic extract was determined according to Refs. [21,22], by mixing 1 mL of phenolic extract which prepared at different concentrations (1, 2.5, and 5 mg/mL) with 2.5 mL of Potassium ferric cyanide solution, then 2.5 mL of phosphate buffer solution at a concentration of 0.2 M at a pH 6.6 was added to the mixture, the mixture was incubated for 20 min at 50 °C then 2.5 mL of 10% Trichloroacetic acid (TCA). A centrifugation step was performed at 3000 rpm for 10 min, then 2.5 mL of distilled water was added to 2.5 mL of phenolic extract solution and 1 mL of 1% of ferric chloride, then the mixture was left for 30 min. Finally, absorbance was measured using a spectrophotometer at wavelength 700 nm. An industrial antioxidant Propyl gallate (PG) was used for comparison in this experiment, the following formula was applied to determine the reducing power. Where As is the absorbance of the sample, Ac is the absorbance of the control.

$$\text{Reducing power\%} = 100 - (\text{As}/\text{Ac}*100)$$

### 2.4. Inhibition efficacy test of β-carotene pigment

The inhibitory efficacy of β-carotene pigment has been tested using the agar disc diffusion method against some bacterial species such as *Staphylococcus aureus*, *Escherichia coli, Pseudomonas aeruginosa*, and *Klebsiella pneumoniae* according to Refs. [23] method by creating 5-mm diameter holes in petri dishes containing 2% nutrient agar which was previously added 0.1 $cm^3$ of bacterial suspension from each species, then the holes were filled with 10 microns of β-carotene pigment at a concentration of 100 mg/mL. The cultures were incubated at 37 °C for 24 h, distilled water and Streptomycin were used as negative and positive controls respectively. Then, the diameter of the inhibition zone was measured using a listed ruler and the results were recorded in (mm) units.

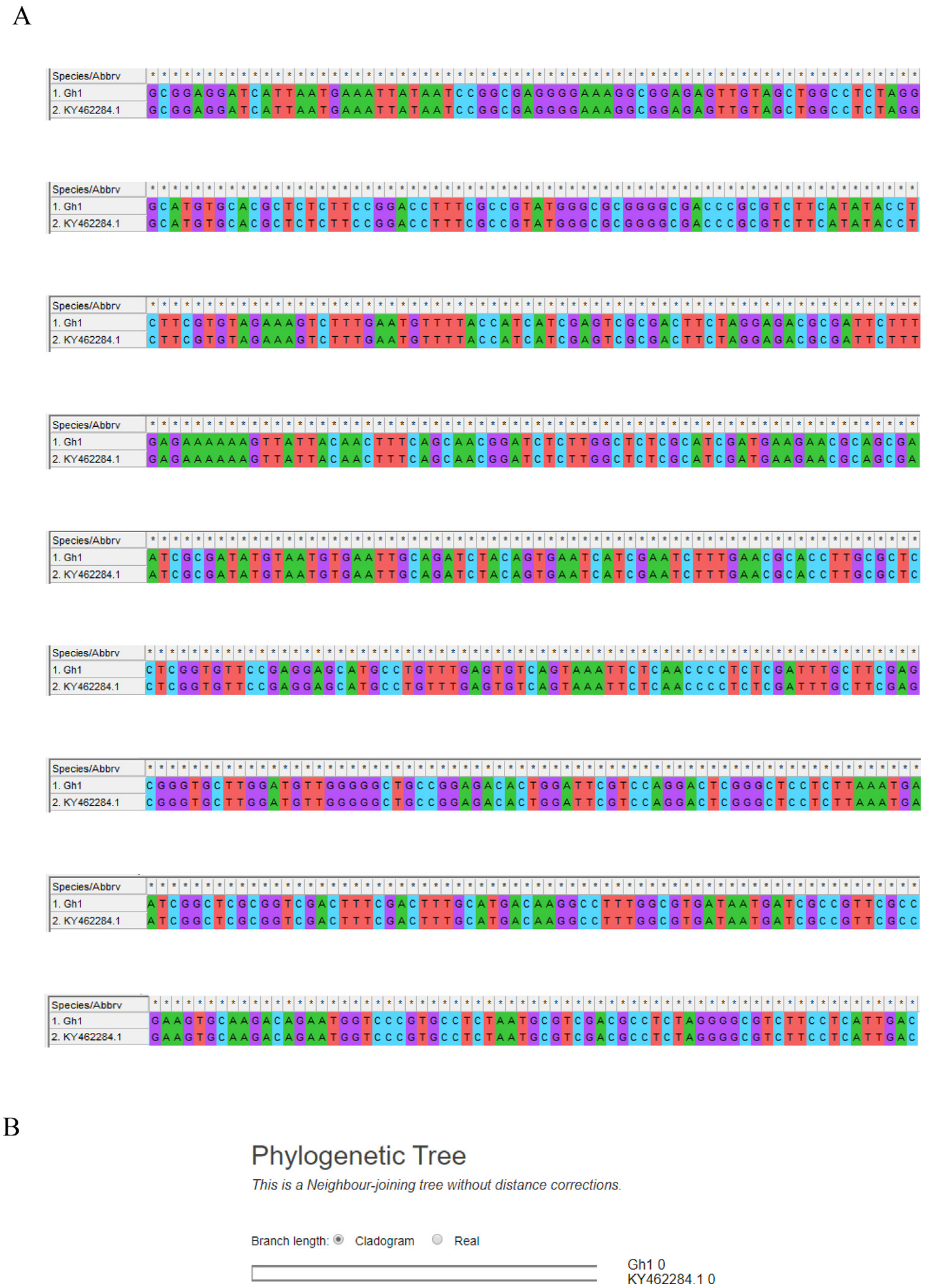

Fig. 2. (A) Multiple alignments of ITS sequences of *Suillus* sp (referred to as Gh1) with deposited sequences in NCBI (Accession: KY462284.1) shows homology of 100% with similar length. (B) Phylogenetic tree, the (Clustal Omega) https://www.ebi.ac.uk/Tools/msa/clustalo/was used to perform the phylogeny analysis, MEGA software has not applied due to only two sequences have been used here.

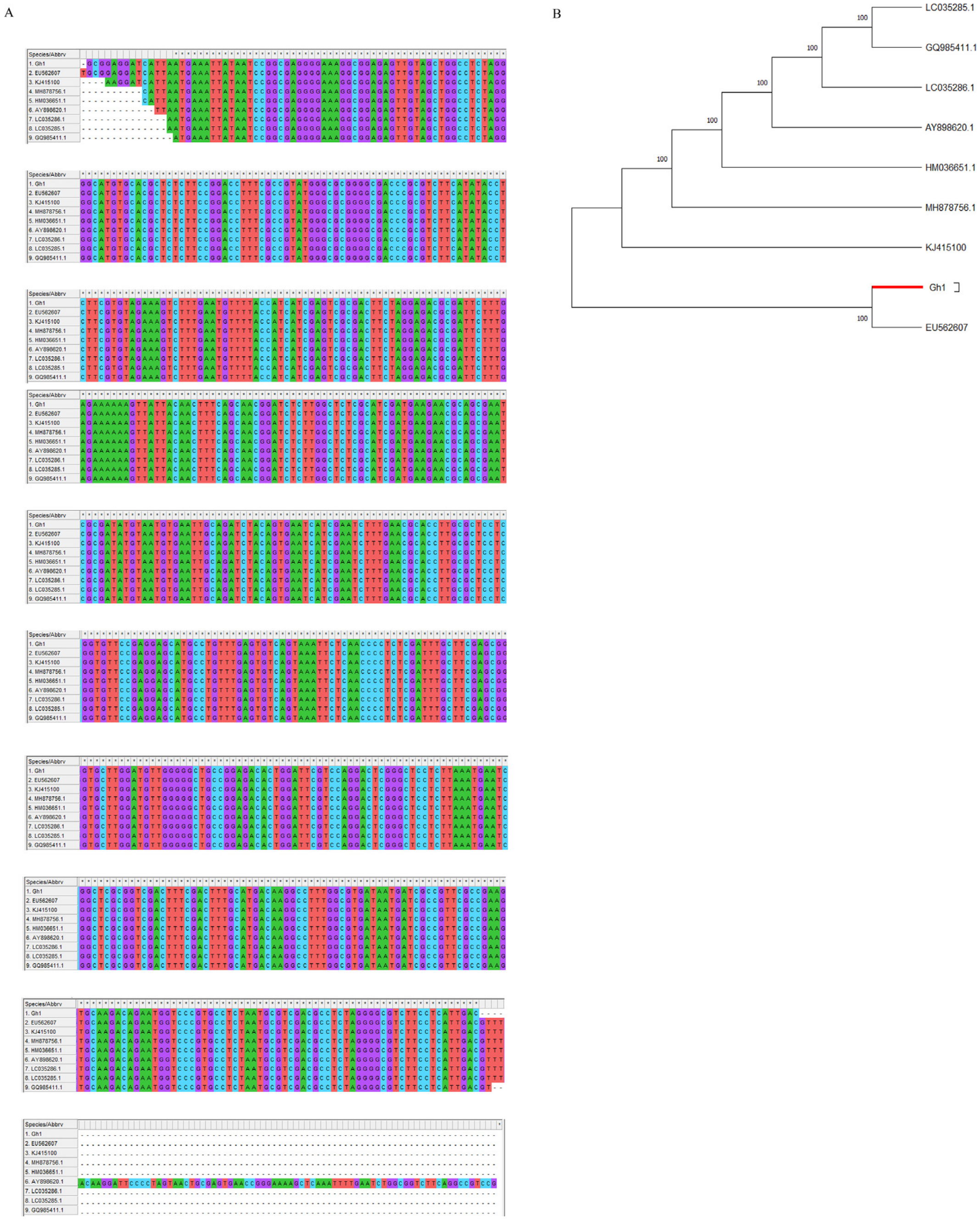

Fig. 3. (A) Multiple alignments of ITS sequences of *Suillus* sp (referred as Gh1) with deposited sequences in NCBI (Accessions: EU562607, KJ415100, MH878756.1, HM036651.1, AY898620.1, LC035286.1, LC035285.1, and GQ985411.1) shows homology of 100% with dissimilar length, (B) Phylogenetic tree: the evolutionary history was inferred using the UPGMA method [25]. The optimal tree with the sum of branch length = 0.00022923 is shown. The percentage of replicate trees in which the associated taxa clustered together in the bootstrap test (100 replicates) are shown next to the branches [26]. Evolutionary distances were computed using the Maximum Composite Likelihood method [27] and are in the units of the number of base substitutions per site. This analysis involved nine nucleotide sequences. Codon positions included were 1st + 2nd + 3rd + Noncoding. All ambiguous positions were removed for each sequence pair (pairwise deletion option). There were a total of 763 positions in the final dataset. Evolutionary analyses were conducted in MEGA X [28].

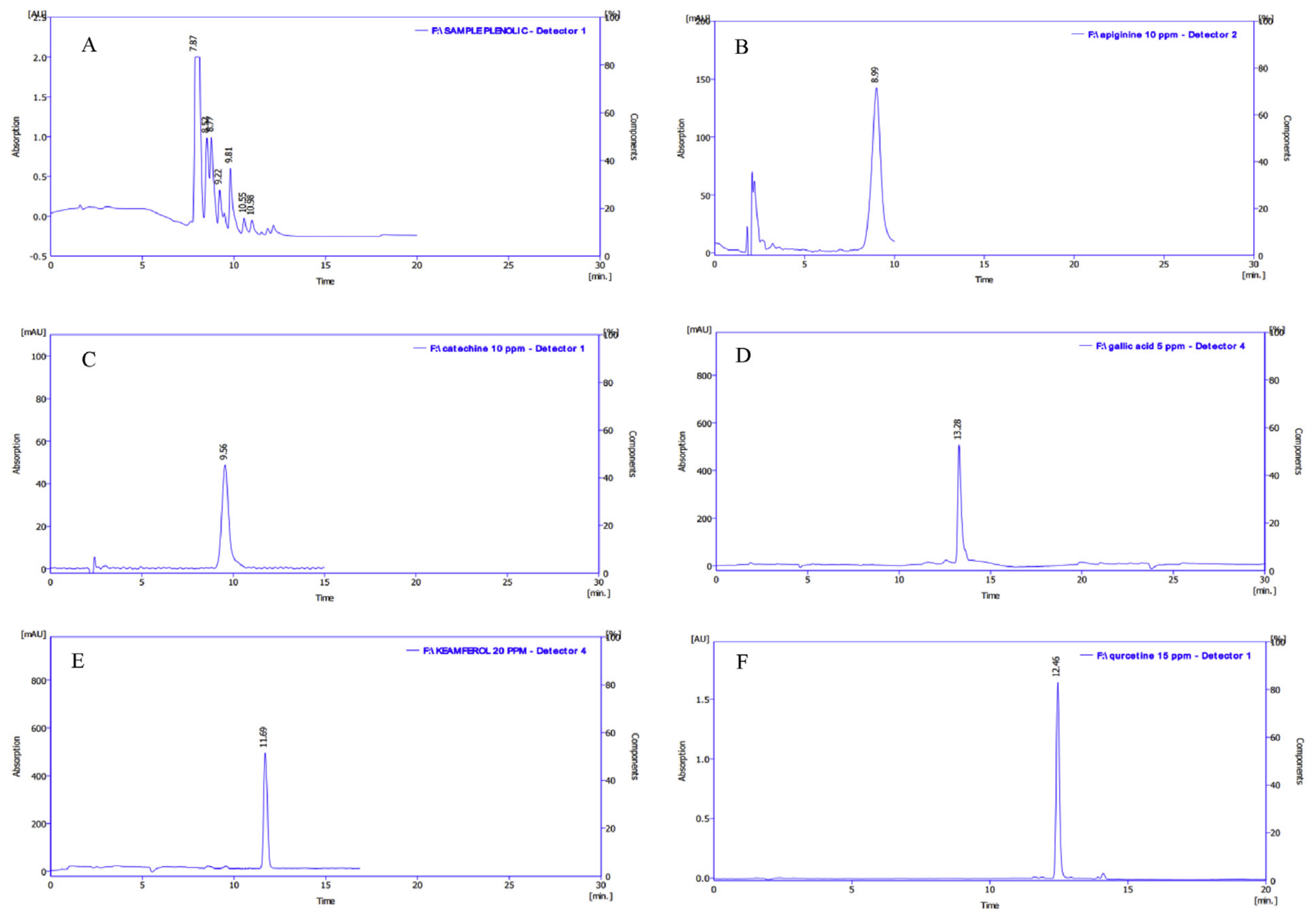

Fig. 4. HPLC profile of (A) all phenolic samples (B) Apiginine (C) Catechine (D) Gallic acid, (E) Keamferol, and (F) Qurcetine. Total run time is 30 min, gradient elution of two solvents were used-methanol and (acetic acid in water). The pigment was detected by (UV–Vis) at a wavelength of 360 nm at a flow rate 1 mL/min.

## 3. Results

### 3.1. DNA extraction and ITS sequence analysis

The DNA from fruiting bodies of *Suillus* sp was extracted using the cetyl Trimethyl Ammonium Bromide (CTAB) method [24], the internal transcribed spacer (ITS) region was amplified by PCR using the universal primers ITS1 (5′-TCCGTAGGTGAACCTGCGG-3′) and ITS4 (5′- TCCTCCGCTTATTGATATGC-3′). The 612-bp fragment was checked by agarose gel electrophoresis in 1% (w/v) bromide and visualized by UV transilluminator (image not shown). The fragment was eluted and purified using Agarose Gel DNA Extraction Kit (Takara, Japan). Five μl primers with 5 μl of purified fragment were sent to Microgen Company/South Korea for sequencing https://dna.macrogen.com/#.

BLAST analysis of the ITS sequence of *Suillus* sp was performed at the BlastN site at the NCBI server https://blast.ncbi.nlm.nih.gov/Blast.cgi?PROGRAM=blastn&PAGE_TYPE=BlastSearch&LINK_LOC=blasthome.

The studied ITS sequence result showed a homology of 100% with equal length of the deposited sequence (Accession: KY462284.1) (Fig. 3), and homology of 100% with the unequal length of deposited sequences (Accessions: EU562607, KJ415100, MH878756.1, HM036651.1, AY898620.1, LC035286.1, LC035285.1, and GQ985411.1). The studied ITS sequence was multaligning with available sequences in NCBI (Figs. 2 and 3), also the evolutionary relationships were performed using MEGA X_10.1.7 software https://www.megasoftware.net/.

### 3.2. HPLC analysis

Culture filtrate (1 mL) of fruiting bodies of *Suillus* sp extracted by ethyl acetate was used for HPLC

Table 1
The retention times of phenolic peaks: all phenolic compounds, Apiginine, Catechine, Gallic acid, Keamferol, and Qurcetine, respectively.

| Result Table (Uncal-F:/Sample phenolic- Detector 1) | | | | | | |
|---|---|---|---|---|---|---|
| S. No | Reten. Time [min] | Area [mAU.s] | Height [mAU] | Area [%] | Height [%] | W 05 [min] |
| 1 | 7.867 | 25608.736 | 1568.328 | 52.5 | 37.2 | 0.34 |
| 2 | 8.520 | 5148.209 | 676.688 | 10.6 | 16.1 | 0.14 |
| 3 | 8.767 | 6066.446 | 615.557 | 12.4 | 14.6 | 0.13 |
| 4 | 9.220 | 2543.595 | 345.414 | 5.2 | 8.2 | 0.12 |
| 5 | 9.807 | 6433.304 | 690.853 | 13.2 | 16.4 | 0.14 |
| 6 | 10.550 | 1619.528 | 173.208 | 3.3 | 4.1 | 0.14 |
| 7 | 10.983 | 1364.330 | 144.018 | 2.8 | 3.4 | 0.15 |
| | Total | 48784.148 | 4214.066 | 100.0 | 100.0 | |
| Result Table (Uncal-F:/Apiginine 10 ppm- Detector 2) | | | | | | |
| 1 | 8.993 | 574.366 | 36.829 | 100.0 | 100.0 | 0.22 |
| | Total | 574.366 | 36.829 | 100.0 | 100.0 | |
| Result Table (Uncal-F:/Catechine 10 ppm- Detector 2) | | | | | | |
| 1 | 9.557 | 288.098 | 20.321 | 100.0 | 100.0 | 0.23 |
| | Total | 574.366 | 20.321 | 100.0 | 100.0 | |
| Result Table (Uncal-F:/Gallic acid 5 ppm- Detector 4) | | | | | | |
| 1 | 9.557 | 795.984 | 128.031 | 100.0 | 100.0 | 0.08 |
| | Total | 795.984 | 128.031 | 100.0 | 100.0 | |
| Result Table (Uncal-F:/Keamferol 20 ppm- Detector 4) | | | | | | |
| 1 | 9.557 | 742. | 92.433 | 100.0 | 100.0 | 0.09 |
| | Total | 795.984 | 92.433 | 100.0 | 100.0 | |
| Result Table (Uncal-F:/qurcetine 15 ppm- Detector 1) | | | | | | |
| 1 | 12.457 | 279.879 | 144.012 | 100.0 | 100.0 | 0.03 |
| | Total | 279.879 | 144.012 | 100.0 | 100.0 | |

analysis according to Ref. [29], this analysis was performed in the laboratories of the Ministry of Science and Technology/Environment and Water Department after the acid hydrolysis process by HPLC model (SYKAM) Germany. The column is C18-ODS and dimensions 4.6 mm* 25 cm. The mobile phase was A,B were A= (methanol: D. W: acetic acid) (85:13:2), B= (methanol: D. W: acetic acid) (25:70:5). Samples were detected at a wavelength of 360 nm at a flow rate of 1 mL per minute. The HPLC analysis showed the presence of five bioactive compounds apiginine, catechine, gallic acid, keamferol and qurcetine (Fig. 4 and Table 1). The amount of $\beta$-carotene pigment was determined by HPLC. The mobile phase was (Acetonitrile: DW) (80: 20), the $\beta$-carotene pigment was detected by (UV—Vis) at a wavelength 450 nm at a flow rate 1 mL/min (Fig. 5 and Table 2).

### 3.3. Antioxidant activity (reducing power)

The (Fig. 6) shows a significant ability of phenolic extract of fruiting bodies of *Suillus* sp to reducing ferric ions ($F^{+3}$) to ferrous ions ($F^{+2}$) comparing with the industrial antioxidant (PG) at all used concentrations. The results revealed that the phenolic compounds extract possesses more reducing power than industrial antioxidants (PG). The percent of reducing power of phenolic compounds was 75.5, 84.9, and 95.7% at concentrations of 1, 2.5 and 5 mg/mL respectively; comparing with PG were 65.9, 81.3, and 93.3 at 1, 2.5, and 5 mg/mL respectively. According to the results, the reduction power increases with increasing concentrations, this results in agreement with [30,31].

### 3.4. Antimicrobial activity of $\beta$-carotene pigment

The $\beta$-carotene pigment crude was extracted from fruiting bodies of *Suillus* sp by methanol solvent and was detected for its antimicrobial activity. The result showed considerable antimicrobial activity against *Klebsiella pneumonia, E. coli* and *S. aureus*, *B. subtilis*, but no activity showed with *P. aeruginosa* (Fig. 7 and Table 3). A concentration of 100 mg/mL of $\beta$-carotene pigment was used to screen the antimicrobial activity, distilled water and chloramphenicol antibiotic were used as a negative and positive control, respectively. The highest bacteria growth inhibition was observed in *K. pneumonia* (40 mm), followed by *E. coli* (36 mm) and *S. aureus* (31 mm), while no antimicrobial activity showed against *P. aeruginosa* (Fig. 7 and Table 3).

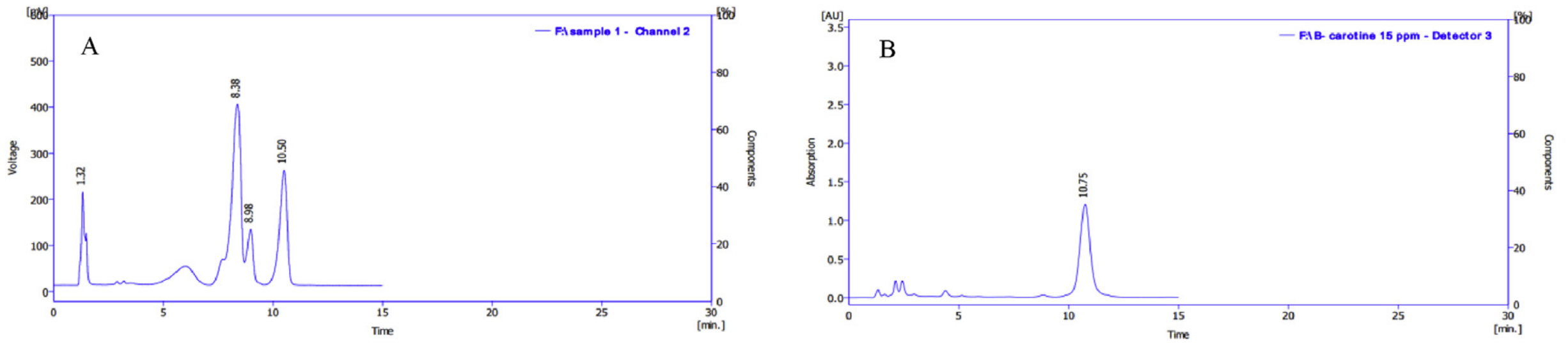

Fig. 5. HPLC profile of all compounds and $\beta$-carotene pigment. Total run time is 30 min, gradient elution of one solvent was used (acetonitrile in water). The pigment was detected by (UV—Vis) at a wavelength of 450 nm at a flow rate 1 mL/min.

Table 2
The retention times of peaks of all compounds and $\beta$-carotene pigment, respectively.

| Result Table (Uncal-F:/Sample 1- Detector 2) | | | | | | |
|---|---|---|---|---|---|---|
| S. No | Reten. Time [min] | Area [mAU.s] | Height [mAU] | Area [%] | Height [%] | W 05 [min] |
| 1 | 1.317 | 1785.066 | 172.238 | 18.4 | 27.0 | 0.12 |
| 2 | 8.383 | 2954.928 | 187.772 | 30.5 | 29.5 | 0.29 |
| 3 | 8.983 | 1003.492 | 75.789 | 10.4 | 11.9 | 0.24 |
| 4 | 10.500 | 3949.246 | 201.699 | 40.7 | 31.6 | 0.32 |
| | Total | 9692.732 | 637.499 | 100.0 | 100.0 | |
| Result Table (Uncal-F:/Carotine 15 ppm- Detector 3) | | | | | | |
| 1 | 10.750 | 2884.671 | 242.315 | 100.0 | 100.0 | 0.19 |
| | Total | 2884.671 | 242.315 | 100.0 | 100.0 | |

## 4. Discussion

The DNA sequence of the ITS region of the collected samples was analyzed to distinguish many species in the genus *Suillus* and other fungi [3,4,32]. The antioxidant properties (reducing power) of methanol extract of the fruiting bodies of *Suillus* sp were analyzed in this study. The elevated antioxidant was determined at a concentration of 5 mg/mL (reducing power = 95.7%), which is higher than industrial antioxidant (PG) (reducing power = 93.3%). The lowest antioxidant was showed at a concentration of 1 mg/mL (reducing power = 75.5%), which is significantly higher than PG (reducing power = 65.9%). This result is similar to Refs. [33] who found a high correlation between anti-oxidative activities and phenolic content concentrations which increased from 23.92%, 36.18% and 79.46% through increasing the concentrations from 100, 500 and 1000 μg/mL respectively. This result inconsistent with [34,35] regarding the relationship between antioxidative activities and phenolic content, the highest antioxidant associated with the greatest amount of phenolic content. Also, the reason for increasing the reducing power of secondary metabolites is probably due to reduced groups that interact with peroxides and prevent their formation [36–39]. This gives an indication of the possibility of using the phenolic extract of *Suillus* sp (bioactive compounds) to be a promising compound as a natural antioxidant which can be applied in the food system, as well as the safety of in inhibiting oxidation by the mechanism of reducing power (antioxidant) which shown in the current study. On the other hand, many studies dealt with the efficacy of a natural β-carotene as an antioxidant, but few have reported the antibacterial activity especially with human pathogenic bacteria and the current paper highlighted this aspect. The inhibitory effect of fruiting bodies of *Suillus* sp methanol extract on some human pathogenic bacteria was evaluated by using the agar well diffusion method [40]. The result revealed a remarkable and greatest antimicrobial activity at 100 mg/mL of β-carotene *K. pneumonia*, followed by *E. coli* and *S. aureus*, *B*, where the inhibition zone diameter was (40, 36, and 31 mm) respectively. However, the *P. aeruginosa* was resistant to the extract and showed no inhibition in growth (Fig. 7 and Table 3). The result is consistent in somehow with [16] that a bovine lactoperoxidase system antimicrobial activity in *S. enteritidis* raised from 1.4 to 3.8 log units by addition of 20-fold diluted CE. β-Carotene, a main natural pigment of carrot.

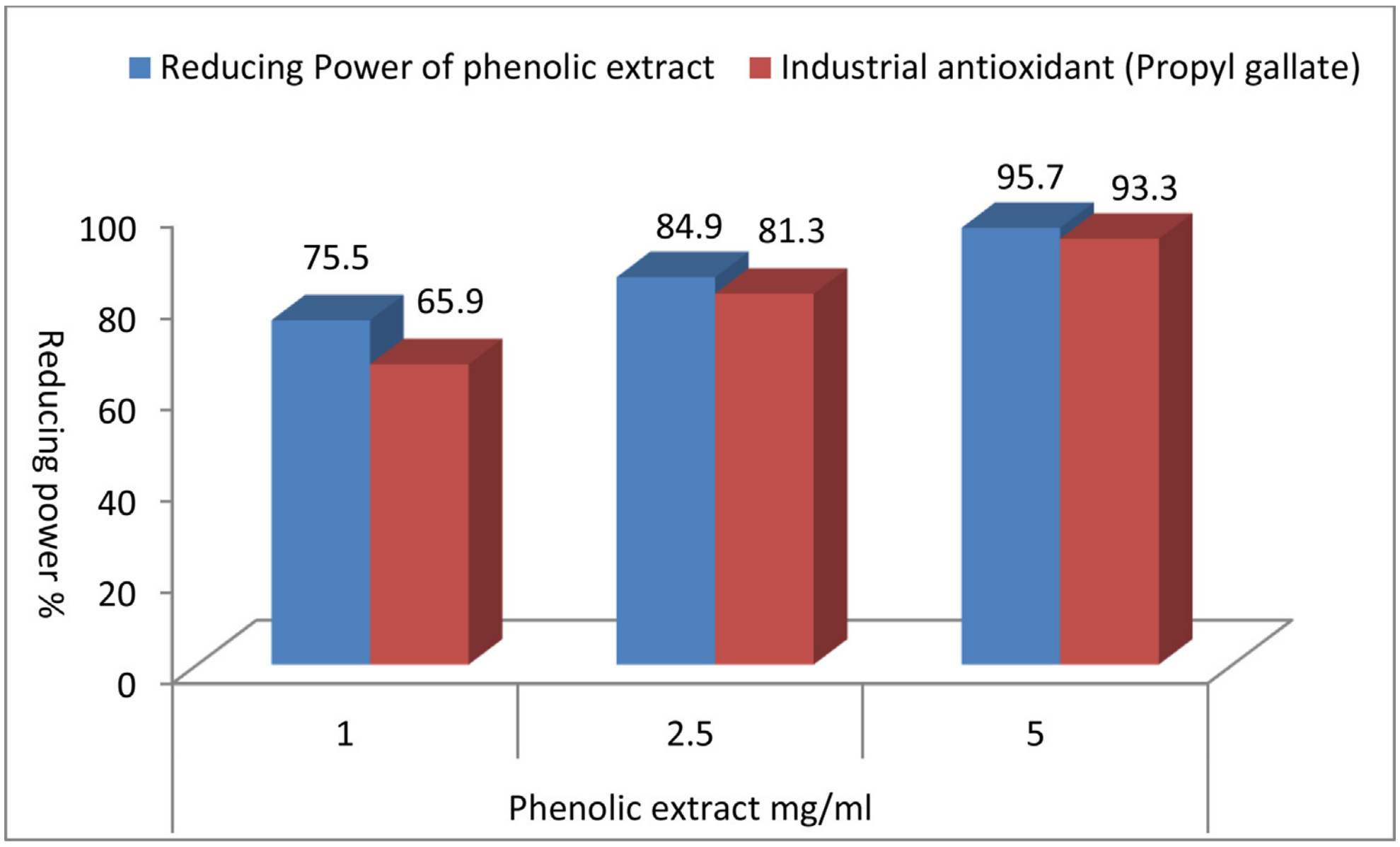

Fig. 6. The average numbers of the percent of reducing power of antioxidant by phenolic extract of *Suillus* sp compared to industrial antioxidant PG at different concentrations.

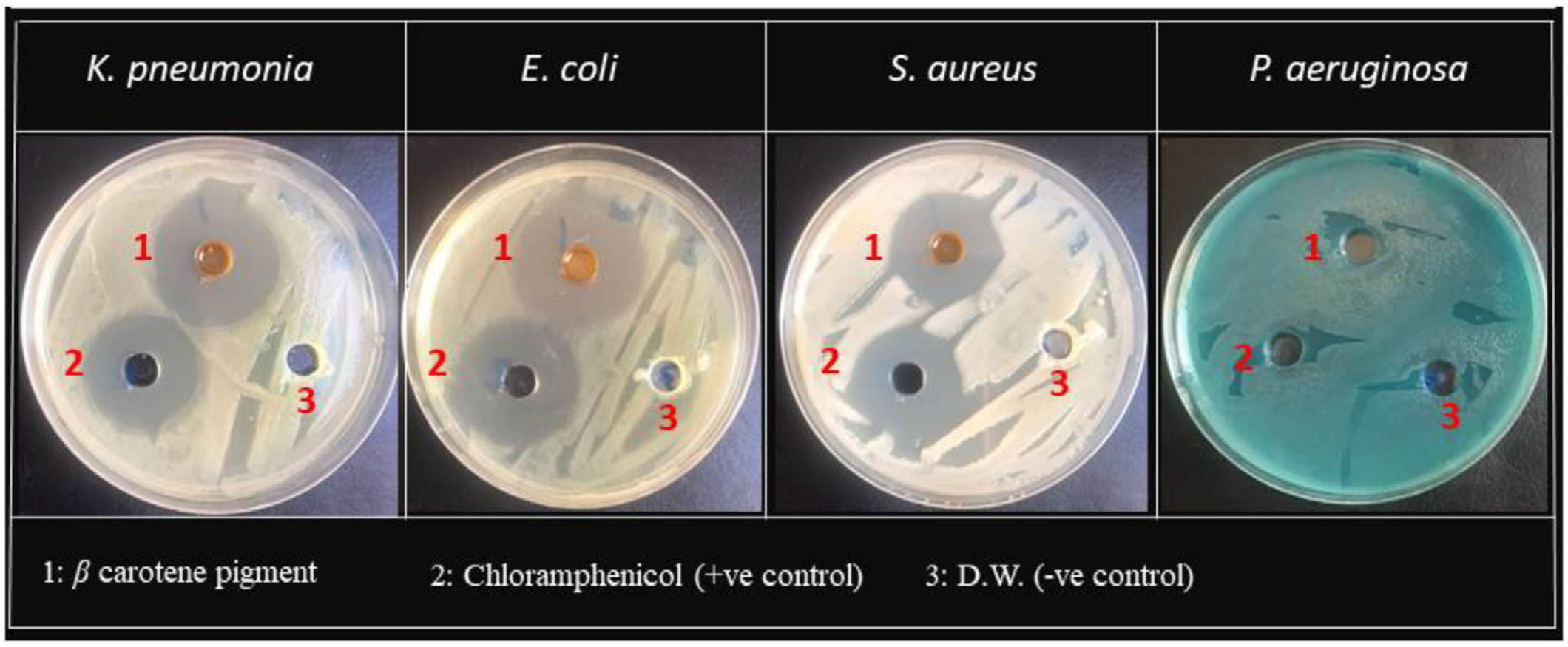

Fig. 7. Antimicrobial activity of methanol extract of β-carotene pigment against bacterial pathogens.

Table 3
The average numbers of antibacterial activity of β-carotene pigment extracted by methanol solvent. Zone of inhibition (Diameter in mm).

| S. No | Microorganisms | Distilled water (−ve control) | Chloramphenicol mg/mL 40 (+ve control) | β-carotene pigment 100 mg/mL |
|---|---|---|---|---|
| 1 | *Klebsiella pneumonia* | 0 | 31 | 40 |
| 2 | *Escherichia coli* | 0 | 32 | 36 |
| 3 | *Staphylococcus aureus* | 0 | 28 | 31 |
| 4 | *Pseudomonas aeuroginosa* | 0 | 0 | 0 |

## 5. Conclusion

To conclude, the molecular identification included the sequencing of the most transcribed region in fungi which is an internal transcribed spacer (ITS) region of (rDNA). The methanol extract of *Suillus* sp a 5 mg/mL showed the highest reducing power (95.7%) comparing with the industrial antioxidant (PG) which was 93.3%, the $\beta$-carotene pigment showed the greatest antimicrobial activity at 100 mg/mL against *K. pneumonia*. The phenolic bioactive compounds isolated from *Suillus* sp can be used as a natural antioxidant which probably having no side effects, also due to the efficacy of antibacterial effects of *β-carotene* as a natural pigment, it could be used as an antibacterial agent to sterilize and disinfect different surfaces, and many other things.

## Acknowledgment

The authors are very grateful to the University of Mosul/College of education for the pure sciences/ Department of Biology for their provided facilities, which helped to improve the quality of this work.